\documentclass[letterpaper]{jpconf}
\usepackage{amsmath}
\newcommand\pp{$\text{\textit{p}} \mathop{\text{+}}
  \text{\textit{p}}$}

\newcommand\CuCu{$\text{Cu} \mathop{\text{+}} \text{Cu}$}

\newcommand\pT{$\text{\textit{p}}_{\text{\textit{T}}}$}

\newcommand\stwohundred{$\sqrt{s} = 200\,\mathrm{GeV}$}

\newcommand\minimize{\mathop{\text{minimize}}}
\newcommand\diag{\mathop{\text{diag}}}
\usepackage{graphicx,amsmath}
\begin{document}
\title{Measurement of jet fragmentation in \pp{} collision at
  $\boldsymbol{\sqrt{s}} \mathop{\text{\textbf{=}}}
  \boldsymbol{200\,}$GeV with the PHENIX detector}

\author{Yue-shi Lai (for the PHENIX Collaboration)}

\address{Columbia University, New York, NY 10027-7061 and Nevis
  Laboratories, Irvington, NY 10533-2508, USA}

\ead{ylai@phys.columbia.edu}

\begin{abstract}
  Measurement of jet fragmentation property in $p + p$ collisions
  provides a crucial baseline for the study of a possible, modified
  fragmentation behavior at the presence of the quark-gluon plasma. We
  describe the first measurement of jet fragmentation functions in
  $\sqrt{s} = 200\,\mathrm{GeV}$ $p + p$ collisions at RHIC. This is
  facilitated by extending the linear least square unfolding technique
  employed by high energy and nuclear experiments to multidimensional
  spectra, which allows us to correct the inefficiencies in the jet
  energy response encountered at detectors such as PHENIX.
\end{abstract}

\section{Introduction}

Hadronization of jet production into final state particles, together
with the production of hard scattering, are complementary parts of the
QCD description of particle production. Fragmentation function in
\pp{} collisions further provides a baseline for any measurement of
the fragmentation property of reconstructed jets in heavy ion
collisions. The narrow detector aperture and the need to operate
within the large multiplicity heavy ion background has traditionally
limited the application of direct jet reconstruction at PHENIX.
Fragmentation properties of jet production therefore been only
measured via two particle correlations by
PHENIX~\cite{Adler:2006sc,Adare:2009vd}. The Gaussian filter jet
reconstruction algorithm~\cite{Lai:2008zp} provides a cone-like
algorithm that at the same time reduces the sensitivity to large angle
fragments (or the lack of such, due to acceptance limit), and is
therefore well-suited for both \pp{} and heavy ion measurement at
PHENIX.

Measurement of fragmentation function using a detector with
inefficiency due to the lack of hadronic calorimetry is difficult,
since the full jet energy is not directly accessible, but has to be
reconstructed from combining track momenta with the electromagnetic
cluster energies. While the energy carried by long-lived neutral
hadrons such as $n$, $K^0_L$ are fully lost, the inherent momentum
resolution and background in a typical tracking system limits the
ability to accurately measure tracks with $p \gg 20\,\mathrm{GeV}/c$.
The resulting large difference in the true and measured jet energy
scale is difficult to correct multiplicatively, but can be addressed
by using unfolding the measured spectra with the
true-to-detector-level jet energy transfer function. This is
particularly difficult for spectra that contain additional dimensions
than the jet energy, such as the fragmentation function. As such, none
of the RHIC experiment, which all lacks hadronic calorimetry, has
attempted the measurement of fragmentation functions so far.

Jet reconstruction using the $\sigma_\mathrm{filter} = 0.3$ Gaussian
filter has been applied to PHENIX for the measurement of the inclusive
jet cross section in \pp{} and jet yield in
\CuCu{}~\cite{Lai:2009zq,Lai:2009ai}. In this proceeding, we report
the development of a multidimensional unfolding technique for the
energy scale correction of fragmentation functions, and the
measurement of \pp{} fragmentation functions for charged particles and
neutral electromagnetic clusters using the PHENIX detector.

\section{Jet reconstruction by Gaussian filtering}

The iterative cone algorithm with the size parameter $R$ is known to
be equivalent to a local maximization of a filter output in $(\eta,
\phi)$ with a special choice of the angular weight function to be the
flat $k(r^2) = \theta(R^2 - r^2)$ and $r^2 = \eta^2 + \phi^2$ (note
that unlike $k(r^2)$, the filter kernel is $h(r^2) \propto -\int dr^2
k(r^2) \propto \max(0, 1 - r^2 / R^2)$, and not
flat)~\cite{Cheng:1995ms,Fashing:2005ms}. Similarly, the Gaussian
filter uses a Gaussian distributed weighting. The Gaussian weighting
takes advantage the property of high-$p_T$ jets being collimated
emissions of particles, and enhances the center signal to the
periphery, which is more likely to be contaminated by the event
background. By avoiding an sharp radial cut-off, the algorithm also
becomes analytically collinear and infrared safe (we further verified
the practical infrared safety using a procedure analogously
to~\cite{Salam:2007xv}).

In \pp{} events and without a large event background, the event
transverse momentum density is the sum of point-like final state
particles $p_{T,i}$
\begin{equation}
  p_T(\eta, \phi) = \sum_{i \in F} p_{T,i} \delta(\eta - \eta_i)
  \delta(\phi - \phi_i).
\end{equation}
The Gaussian filtering of \pT{} is the linear-circular convolution of
$p_T(\eta, \phi)$ with a Gaussian distribution
\begin{equation}
  p_T^\mathrm{filt}\!(\eta, \phi) = \iint d\eta' d\phi' p_T(\eta',
  \phi') \exp\left( -\frac{(\eta - \eta')^2 + (\phi -
      \phi')^2}{2\sigma_\mathrm{filter}} \right).
\end{equation}
The output of the filter for a given $(\eta, \phi)$ position is the
Gaussian weighted transverse momentum in the said event. The local
maxima in $p_T^\mathrm{filt}(\eta, \phi)$ are the reconstructed jets
using the Gaussian filter. More detailed discussion of the property of
Gaussian filter for jet reconstruction can be found
in~\cite{Lai:2008zp,Lai:2009ai}.

\section{Multidimensional unfolding}

The true particle level and the measured energy scale in the PHENIX
central arm differ due to inefficiencies, the lack of hadronic
calorimetry, and the loss of out-of-acceptance fragments due to the
$\Delta\eta = 0.7$ pseudorapidity coverage and also the $\Delta\phi =
\pi$ partial azimuthal coverage. This difference means that a full
unfolding is needed to extract the true energy instead of a first
order multiplicative correction.

Linear least square unfolding with the Phillips--Tikhonov
regularization~\cite{Phillips:1962tf,Tychonoff:1963or} has been widely
used by e.g.\ \textsc{guru}~\cite{Hoecker:1996sa} for the unfolding of
spectra. Binned spectra can be mathematically regarded as a vector of
individual bins, and the linear bin-to-bin migration due to differing
energy scales is then conveniently expressed as a matrix. For a
typical transfer matrix $\mathbf{A}$ generated using Monte Carlo
detector simulation, if the measured spectrum $\mathbf{b}$ with the
covariance matrix $\mathbf{C}$ is unfolded using the simple linear
least square relation
\begin{equation}
  \minimize_\mathbf{x} \quad \lVert \mathbf{A} \mathbf{x} - \mathbf{b}
  \rVert_\mathbf{C}^2,
\end{equation}
($\lVert \mathbf{u} \rVert_\mathbf{C} = \mathbf{u}^T \mathbf{C}^{-1}
\mathbf{u}$ denotes the Mahalanobis distance) the propagation of
fluctuations in $\mathbf{A}$ would result in large, non-statistical
fluctuation in the unfolding result $\mathbf{x}$. The regularized
unfolding determines the unfolded spectrum $\mathbf{x}$ by solving the
minimization problem
\begin{equation}
  \minimize_\mathbf{x} \quad \lVert \mathbf{A} \mathbf{x} - \mathbf{b}
  \rVert_\mathbf{C}^2 + \tau \lVert \mathbf{L} \mathbf{x} \rVert^2,
  \label{eq:regularized_linear_least_square}
\end{equation}
where $\tau$ is the regularization parameter and $\mathbf{L}$ is a linear
operator that describes the amount of discontinuity or ``noisiness''
of the unfolding result.

Spectra in particle and nuclear physics can cross multiple orders of
magnitude and a homogeneous $\mathbf{L}$ would cause the large
magnitude part of the spectra to dominate the singular vectors, thus
exhausting most degrees of freedom purely to reproduce the variation
in magnitudes. This can be solved by either scaling $\mathbf{L}$ to
approximately match the variations in $\mathbf{x}$, or to prescale
$\mathbf{x}$. We follow the approach used by \textsc{guru} to prescale
the unknowns by
\begin{equation}
  x_i \mapsto \hat{x}_i = \frac{x_i}{x_i^\mathrm{ini}},
\end{equation}
and the linear system $\mathbf{A}$ is therefore inversely scaled by
\begin{equation}
  a_{ij} \mapsto \hat{a}_{ij} = a_{ij} x_i^\mathrm{ini}
\end{equation}
(column-wise scaling), with the purpose of preventing exceedingly
large or small $\hat{x}_i$ to appear numerically.

The typical choice for $\mathbf{L}$ is a second order
finite-difference matrix
\begin{equation}
  \mathbf{L}^{(2)} = \begin{pmatrix}
      -1 & 1 & 0 & \cdots & 0\\
      1 & -2 & 1 & & \vdots\\
      0 & 1 & -2 & & 0\\
      \vdots & & & \ddots & 1\\
      0 & \cdots & 0 & 1 & -1
  \end{pmatrix}
\end{equation}
The minimization with respect to the second order derivative describes
a continuity constraint that restricts the shape of $\hat{\mathbf{x}}$
to be cubic spline-like. The simplest $D$-dimensional generalization
of $\mathbf{L}^{(2)}$ are the isotropic axial derivatives
$\partial^2/\partial x_k^2, k = 1, \dots, D$, and the minimization
with respect to it behaves similarly to the $D$-dimensional cubic
tensor splines. We found this type of regularization to be sufficient
for the purpose of unfolding multidimensional spectra with Poisson
statistics, other choices used e.g.\ in image restoration are reviewed
in \cite{Karayiannis:1990rt}.

The solution to \eqref{eq:regularized_linear_least_square} can be
written as a least square problem with the matrix pencil
$(\hat{\mathbf{A}} + \sqrt{\tau} \mathbf{L})$, and can be solved by a
SVD of $(\boldsymbol{\Lambda}^{-1} \hat{\mathbf{A}}) \mathbf{L}^{-1} =
\mathbf{U} \boldsymbol{\Sigma} \mathbf{V}^T$, where $\mathbf{C} =
\boldsymbol{\Lambda} \boldsymbol{\Lambda}^T$ is the Cholesky
decomposition of the measurement covariance, and provided $\mathbf{L}$
is nonsingular. The software package \textsc{guru} provides an
implementation of this method for diagonal $\mathbf{C}$ and using
$\mathbf{L}^{(2)}$ (which is strictly speaking singular, but the
rectangular nature allows it to be approximately inverted by
perturbing $\mathbf{L}^{(2)} \mapsto \mathbf{L}^{(2)} + \epsilon
\mathbf{I}$, $\lvert\epsilon\rvert \ll 1$). However, a $D$-dimensional
continuity constraint for $N$ measurements consists of at least $DN$
axial derivatives and therefore $\mathbf{L}$ becomes a $DN\times N$
matrix, it is therefore generally not possible to invert $\mathbf{L}$
and solve the multidimensional unfolding in the fashion of
\textsc{guru}.

The least square problem with a matrix pencil
$(\boldsymbol{\Lambda}^{-1} \hat{\mathbf{A}} + \sqrt{\tau}
\mathbf{L})$ can also be solved by the generalized SVD
(GSVD)~\cite{VanLoan:1976gt} of a matrix pair
$\mathrm{GSVD}(\boldsymbol{\Lambda}^{-1} \hat{\mathbf{A}},
\mathbf{L})$, which simultaneously decomposes both matrices into
\begin{equation}
  \begin{split}
    \boldsymbol{\Lambda}^{-1} \hat{\mathbf{A}} &= \mathbf{U}
    \begin{pmatrix}
      0 & \boldsymbol{\Sigma}_1
    \end{pmatrix} \mathbf{X}^{-1}\\
    \mathbf{L} &= \mathbf{V}
    \begin{pmatrix}
      0 & \boldsymbol{\Sigma}_2
    \end{pmatrix} \mathbf{X}^{-1}.
  \end{split}
\end{equation}
$\mathrm{GSVD}(\boldsymbol{\Lambda}^{-1} \hat{\mathbf{A}},
\mathbf{L})$ is therefore closely related to
$\mathrm{SVD}((\boldsymbol{\Lambda}^{-1} \hat{\mathbf{A}})
\mathbf{L}^{-1})$, which can be immediately seen by comparing the GSVD
of the quotient matrix $(\boldsymbol{\Lambda}^{-1} \hat{\mathbf{A}})
\mathbf{L}^\dagger = \mathbf{U} (\boldsymbol{\Sigma}_1
\boldsymbol{\Sigma}_2^{-1}) \mathbf{V}^T$ ($\mathbf{L}^\dagger$ being
the $(\boldsymbol{\Lambda}^{-1} \hat{\mathbf{A}})$-weighted
pseudoinverse of $\mathbf{L}$~\cite{Elden:1982wp}) to the ordinary SVD
form $(\boldsymbol{\Lambda}^{-1} \hat{\mathbf{A}}) \mathbf{L}^{-1} =
\mathbf{U} \boldsymbol{\Sigma} \mathbf{V}^T$.

With the generalized singular value pairs $\boldsymbol{\Sigma}_1 =
\diag(\alpha_i)$, $\boldsymbol{\Sigma}_2 = \diag(\beta_i)$, the SVD
singular values of $(\boldsymbol{\Lambda}^{-1} \hat{\mathbf{A}})
\mathbf{L}^\dagger$ are $\gamma_i = \alpha_i/\beta_i$. The solution
then can be written as the Tikhonov filter for the singular values
\begin{equation}
  {\boldsymbol{\Sigma}'}^{-1} = \diag \left(
    \frac{\gamma_i^2}{\gamma_i^2 + \tau} \alpha_i^{-1} \right),
\end{equation}
and the mean and covariance matrix of the unfolding solution is
\begin{equation}
  \begin{split}
    \hat{\mathbf{x}} &= \mathbf{X} {\boldsymbol{\Sigma}'}^{-1}
    \mathbf{U}^T (\boldsymbol{\Lambda}^{-1} \mathbf{b})\\
    \mathop{\mathrm{Cov}} \hat{\mathbf{x}} &= \mathbf{X}
    {\boldsymbol{\Sigma}'}^{-1} ({\boldsymbol{\Sigma}'}^{-1})^T
    \mathbf{X}^T.
  \end{split}
\end{equation}

We implemented this multidimensional generalization to \textsc{guru}'s
regularized linear least square unfolding based on \textsc{lapack}'s
\textsc{dggsvd} \cite{Anderson:1999lu}, which implements the dense GSVD
algorithm by Bai, Demmel \& Zha \cite{Bai:1993np,Bai:1993ct}.

\section{Experimental setup}

\begin{figure}[t]
  \begin{minipage}[t]{3in}
    \centerline{\includegraphics[width=2.75in]{%
        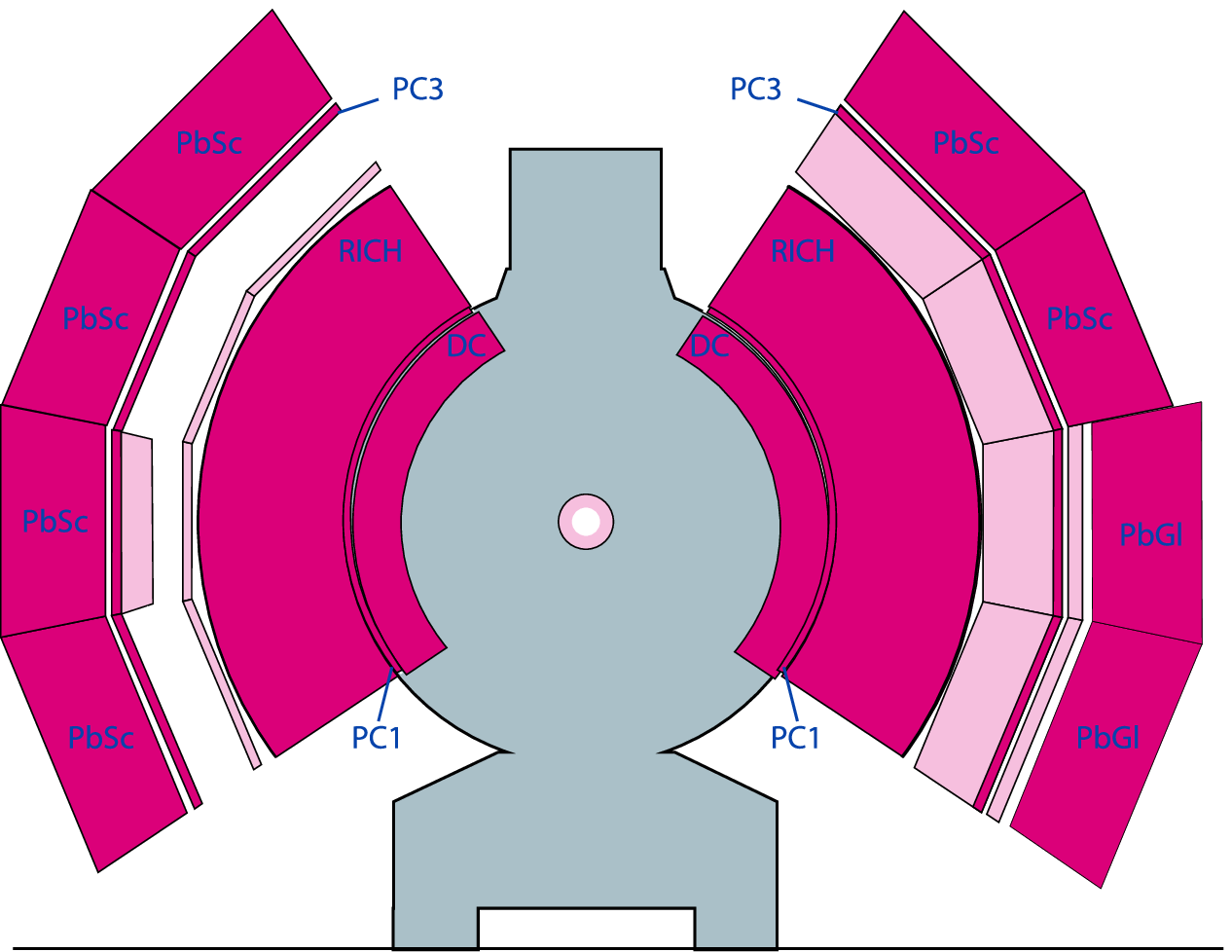}}
    \caption{The PHENIX central arm detectors for RHIC Run-5 (year
      2004/2005), viewed along the beam axis from the south towards
      north. Dark regions indicate detectors used for the jet
      reconstruction: The drift chamber (DC), the pad chamber layers 1
      and 3 (PC1/PC3), the ring-imaging \v{C}erenkov detector (RICH),
      and the Pb scintillator (PbSc) and Pb glass (PbGl) electromagnetic
      calorimeters.}
    \label{fig:central_arm_run_4_jet_reconstruction}
  \end{minipage}\hfill%
  \begin{minipage}[t]{3in}
  \centerline{\includegraphics[width=3in]{%
      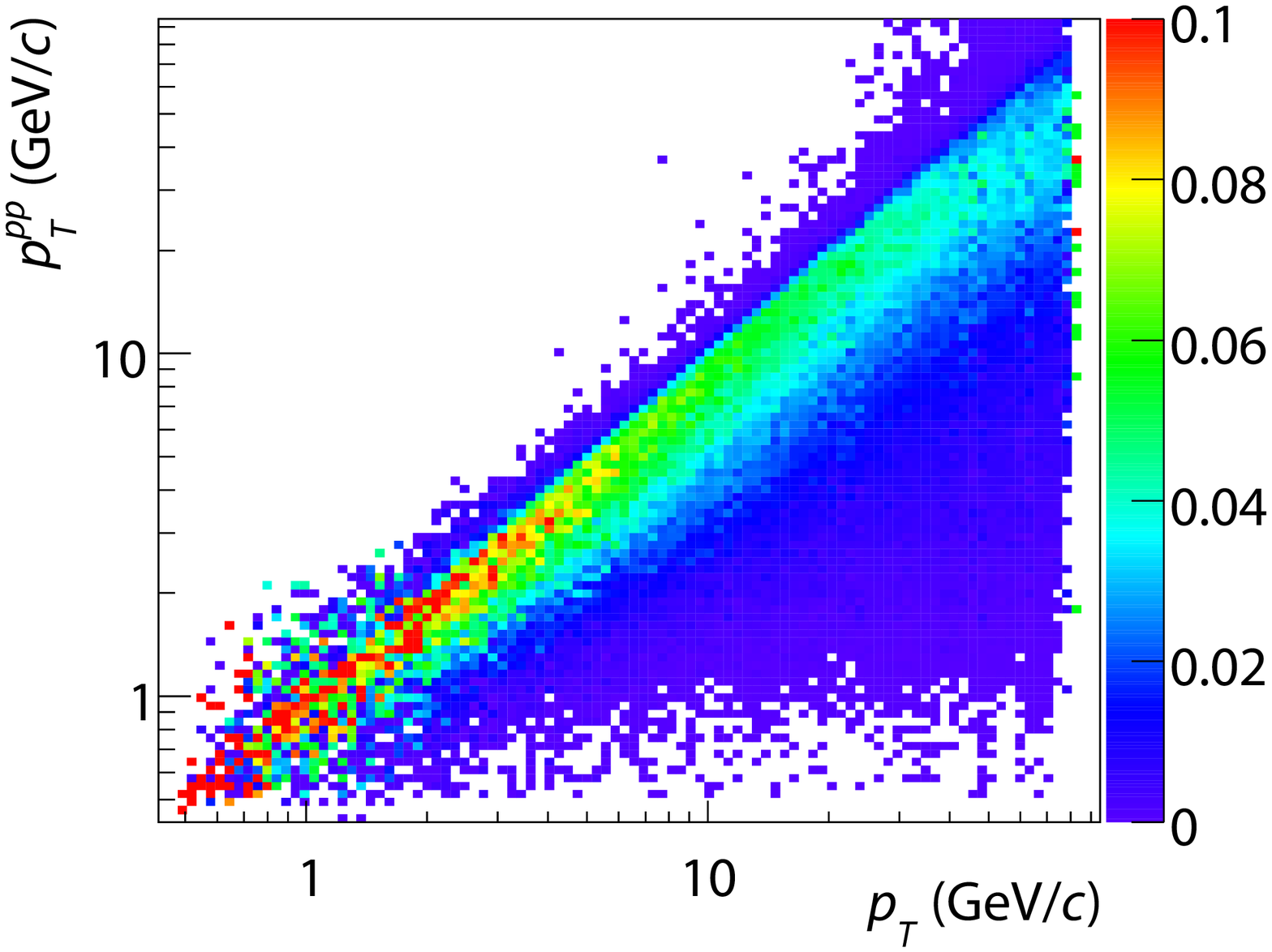}}
  \caption{The PHENIX jet $P(p_T^{pp} | p_T)$ transfer matrix for
    \stwohundred{} and $\sigma = 0.3$ Gaussian filter, derived from
    the \textsc{geant} simulation of $\approx 1.6\times 10^7$
    \textsc{pythia} events. The $p_T^{pp} < p_T$ region is dominated
    by $n$, $K_L^0$ energy loss.}
  \label{fig:p_p_energy_scale}
  \end{minipage}
\end{figure}

Figure \ref{fig:central_arm_run_4_jet_reconstruction} shows the Run-5
PHENIX ``central arm'' configuration for RHIC Run-5 (year 2004/2005).
PHENIX has two mid-rapidity spectrometers with an aperture of $|\eta|
< 0.35$ and $\Delta\phi = \pi/2$ each. The components used for jet
reconstruction are the drift chamber (DC), the pixel pad chamber
layers 1 and 3 (PC1/PC3), the ring-imaging \v{C}erenkov detector
(RICH), and the electromagnetic calorimeters (EMCal). For the data
presented in this paper, DC/PC1/PC3 provide momentum measurement for
charged particles, and the EMCal the energy for photons.

Pattern recognition and momentum reconstruction in the tracking system
formed by DC and PC1/3 is performed using the combinatorial Hough
transform, with the momentum scale determined by the time-of-flight
measurement of identified $\pi^\pm$, $K^\pm$, and $p$/$\bar{p}$. The
momentum resolution of the tracking system is determined as $\delta
p/p = 0.7\% \oplus 1.0\% p / (\mathrm{GeV}/c)$. Two calorimeter
technologies were used, six of the total eight sectors are covered by
Pb-scintillator (PbSc) calorimeters with resolution of $\sigma_E/E =
8.1\%/\sqrt{E} \oplus 2.1\%$ and a granularity of $\Delta\eta \times
\Delta\phi \approx 0.01\times 0.01$, and two sectors by Pb-glass
(PbGl) calorimeters with $\sigma_E/E = 5.9\%/\sqrt{E} \oplus 0.8\%$
and $\Delta\eta \times \Delta\phi \approx 0.008\times 0.008$. The
intrinsic timing resolution for $1\,\mathrm{GeV}$ $\pi^\pm$ are about
$200$--$300\,\mathrm{ps}$ for both technologies.

Since PHENIX currently lacks an in-field tracking capability,
conversion electrons in the DC can produce a displaced track that has
the appearance a high $p_T$ track originating from the event vertex.
We use the information from the RICH and $dE/dx$ measurement to
identify and remove these tracks. To provide additional suppression at
the cross section level of jets with $p_T > 20\,\mathrm{GeV}$, we use
the fact that conversion electrons are geometrically unlikely to
coincide with the direction of the jet production, and require the
reconstructed jet to have a minimum multiplicity of $3$ particles
measured within a radial angle of $60^\circ$, and the charged fraction
of the jet $p_T$ to be below $0.9$ to remove single track events, or
when the jet is dominated by a large $p_T$ track.

The absolute energy scale of the calorimeter clusters are determined
using both the reconstructed $\pi^0$ masses from the observed $\pi^0
\rightarrow \gamma\gamma$ decays, and checked using the $E/p$ from
energies from cluster matching RICH identified $e^\pm$ tracks. Shower
shape cuts are used to remove clusters. The residual uncertainty in
the energy scale is $\pm 3\%\text{ (syst.)}$. Since the measurement
extends to very low cross sections, processes such as upward beam
interaction can deposit energy in the EMCal that randomly coincides
with an event. This is suppressed by measuring the time-of-flight of
the clusters and rejecting those that are out of synchronization with
the collision.

The PHENIX minimum bias (MB) trigger is defined by the coincident
firing of the two beam-beam counters (BBC) located at $3.0 < \eta <
3.9$. The Van de Meer/vernier scan method is used to measure the BBC
cross section, with $\sigma_\mathrm{BBC} = 22.9\pm
2.3\,\mathrm{mb}\text{ (syst.)}$. The efficiency of BBC MB trigger on
an event containing a jet with $p_T^\mathrm{rec} > 2\,\mathrm{GeV}/c$
is $\epsilon_\mathrm{BBC} = 0.86\pm 0.05\text{ (syst.)}$ and constant
with respect to $p_T^\mathrm{rec}$ within that uncertainty. We require
the collision vertex to be within $|z| < 25\,\mathrm{cm}$ along the
beam axis, derived from the timing difference between the firing of
the two BBC.

\section{Inclusive jet cross section in \pp{}}

\begin{figure}[t]
  \begin{minipage}[t]{3in}
    \centerline{\includegraphics[width=3in]{%
        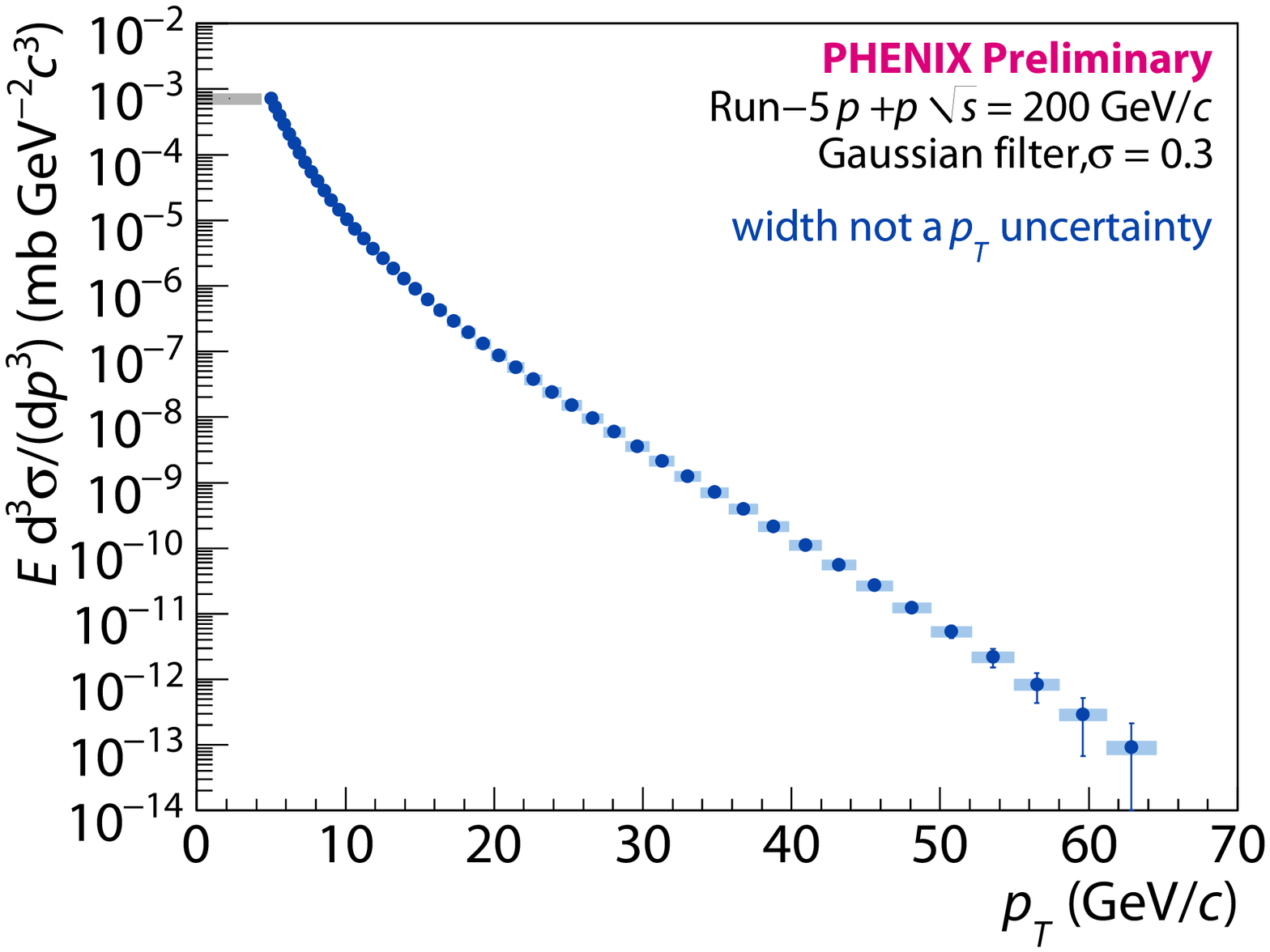}}
    \caption{PHENIX Run-5 \pp{} at \stwohundred{} invariant jet cross
      section spectrum as a function of $p_T$. The shaded box to the
      left indicates the overall normalization systematic uncertainty,
      shaded boxes associated with data points indicate point-to-point
      systematic uncertainties, and error bars indicate statistical
      uncertainties.}
    \label{fig:run_5_p_p_perp}
  \end{minipage}\hfill%
  \begin{minipage}[t]{3in}
    \centerline{\includegraphics[width=3in]{%
        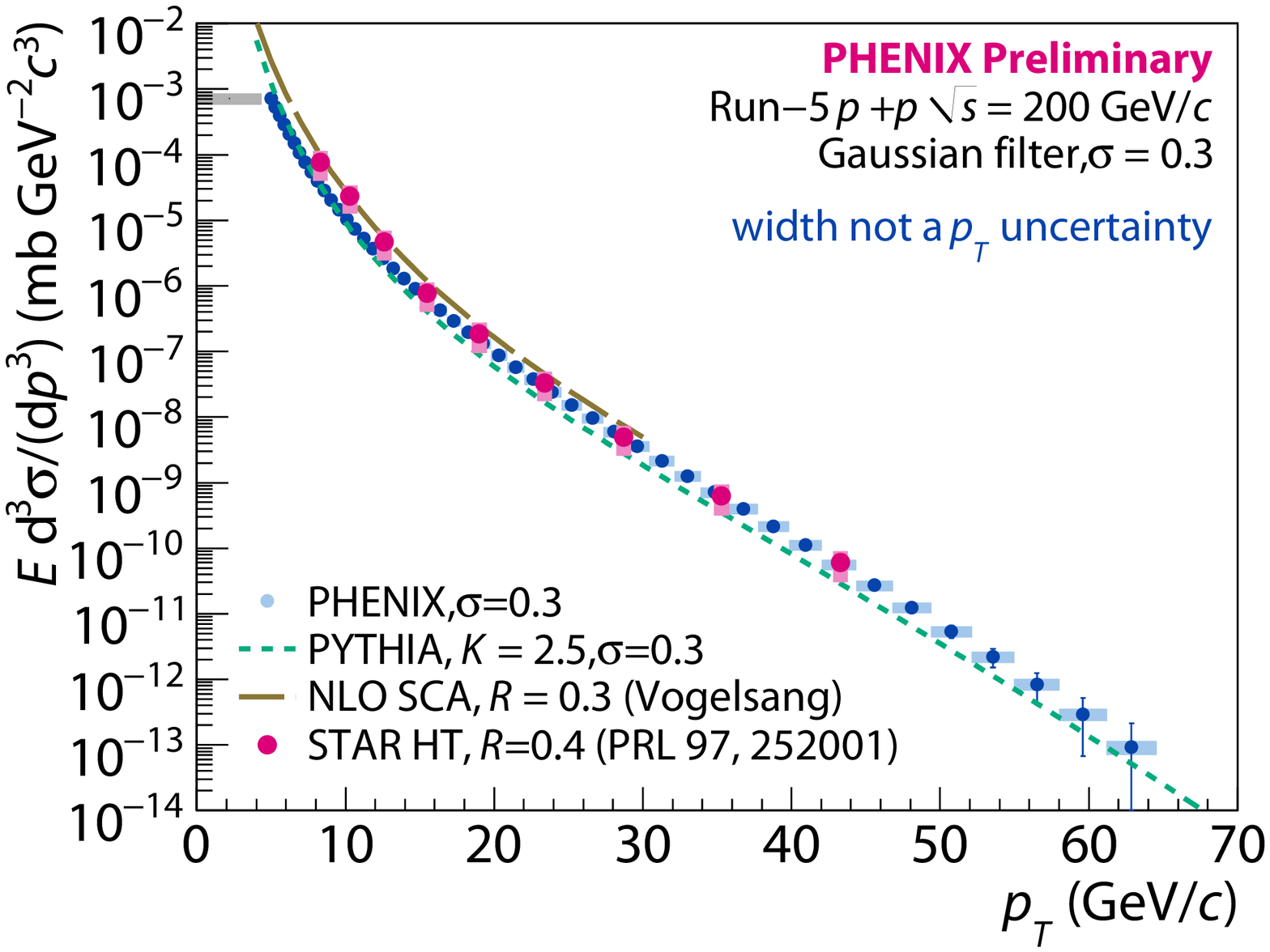}}
    \caption{PHENIX Run-5 \pp{} at \stwohundred{} invariant jet cross
      section spectrum as a function of $p_T$, with comparison to
      \cite{Abelev:2006uq}, next-to-leading order calculation from
      \cite{Jager:2004jh}, and \textsc{pythia} assuming $K = 2.5$. The
      shaded box to the left indicates the overall normalization
      systematic uncertainty, shaded boxes associated with data points
      indicate point-to-point systematic uncertainties, and error bars
      indicate statistical uncertainties.}
    \label{fig:run_5_p_p_perp_comparison_all}
  \end{minipage}
\end{figure}

The data presented in the following sections were obtained from the
PHENIX \pp{} dataset from the RHIC Run-5 (year 2004/2005). After
removal of bad quality runs, a total of $1.47\times 10^9$ minimum bias
\pp{} and $1.16\times 10^9$ triggered \pp{} events are being used.

The measurement of inclusive jet cross section in Run-5 \pp{}
collisions uses the combined minimum bias and triggered data set, and
is based on the regularized unfolding of the jet spectrum
$dN_\mathrm{jet}/dp_T^\mathrm{jet,rec}$ by the jet energy scale
transfer function $P(p_T^\mathrm{jet,rec} | p_T^\mathrm{jet})$, which
is evaluated by \textsc{geant} simulation using \textsc{pythia} 6.4.20
events. A total of $1.6\times 10^7$ events were simulated with 14
different minimum $Q^2$ settings varying between$\sqrt{Q^2} >
0.5\,\mathrm{GeV}/c$ and $\sqrt{Q^2} > 64\,\mathrm{GeV}/c$. The
resulting transfer matrix $P(p_T^{pp} | p_T)$ is shown in figure
\ref{fig:p_p_energy_scale}.

The unfolding is equivalent to the inversion of the 1D Fredholm
equation
\begin{equation}
  \frac{dN_\mathrm{jet}}{dp_T^\mathrm{jet,rec}} = \int
  dp_T^\mathrm{jet} \, P(p_T^\mathrm{jet,rec} | p_T^\mathrm{jet})
  \frac{dN}{dp_T^\mathrm{jet}}.
  \label{eq:fredholm_1d}
\end{equation}
Then the invariant cross section can be evaluated as
\begin{equation}
  \frac{E d^3\sigma^\mathrm{jet}}{dp^3} = \frac{1}{2\pi p_T}
  \frac{d^2\sigma^\mathrm{jet}}{dp_T^\mathrm{jet} dy} =
  \frac{\sigma_\mathrm{BBC}}{A\,\epsilon_\mathrm{BBC}}
  \frac{1}{p_T^\mathrm{jet}} \frac{1}{N_\mathrm{evt}}
  \frac{dN^\mathrm{jet}}{dp_T^\mathrm{jet}}
  \label{eq:sigma_pp}
\end{equation}
where $dN/dp_T$ is the unfolding result from \eqref{eq:fredholm_1d}
and
\begin{equation}
  A = 2 (\Delta\eta - 2 d) (\Delta\phi / 2 - 2 d)
  \label{eq:fiducial_area}
\end{equation}
the fiducially reduced PHENIX central arm acceptance area. This
measurement is detailed in \cite{Lai:2009ai}.

The choice of regularization parameter $\tau$ can translate into an
uncertainty on the low frequency, global shape of a spectrum. We
address this by evaluating the systematic uncertainty from varying
$\tau$ over the entire meaningful range between $\approx 4^D$ degrees
of freedom up to the Nyquist frequency, $D$ being the unfolding
dimension.

Figure~\ref{fig:run_5_p_p_perp} shows the PHENIX preliminary \pp{} jet
spectrum measured using the Gaussian filter, plotted in invariant
cross sections. The shaded box to the left indicates the overall
normalization systematic uncertainty, shaded boxes associated with
data points indicate point-to-point systematic uncertainties, and
error bars indicate statistical uncertainties. We show the unfolded
spectrum out to the $p_T$ bin where the nominal yield for the number
of sampled events reaches the level of $1$ jet, namely
$60\,\mathrm{GeV}/c$.

Figure~\ref{fig:run_5_p_p_perp_comparison_all} shows the same spectrum
as in Figure~\ref{fig:run_5_p_p_perp}, compared against the spectrum
from \cite{Abelev:2006uq}, the next-to-leading order (NLO) calculation
using the small cone approximation (SCA)~\cite{Jager:2004jh}, and the
leading order \textsc{pythia} spectrum assuming $K = 2.5$. The
comparison to \cite{Abelev:2006uq} and NLO SCA involve different jet
definitions, a residual difference should be expected, even though for
$p_T > 15\,\mathrm{GeV}/c$ it appears to be small between filter and
cone jets for the Gaussian size $\sigma = 0.3$ used in this analysis.
Our spectrum is close to \cite{Abelev:2006uq} within its \pT{} reach.
The spectrum also follows approximately the shape of the NLO SCA
calculation, and the leading order \textsc{pythia} spectrum, if $K =
2.5$ is assumed. However, a more appropriate comparison would involve
Gaussian filter based NLO calculations, which we plan to perform in
the future.

\section{Fragmentation functions in \pp{}}

\begin{figure}[t]
  \begin{minipage}[t]{3in}
    \centerline{\includegraphics[width=3in]{%
        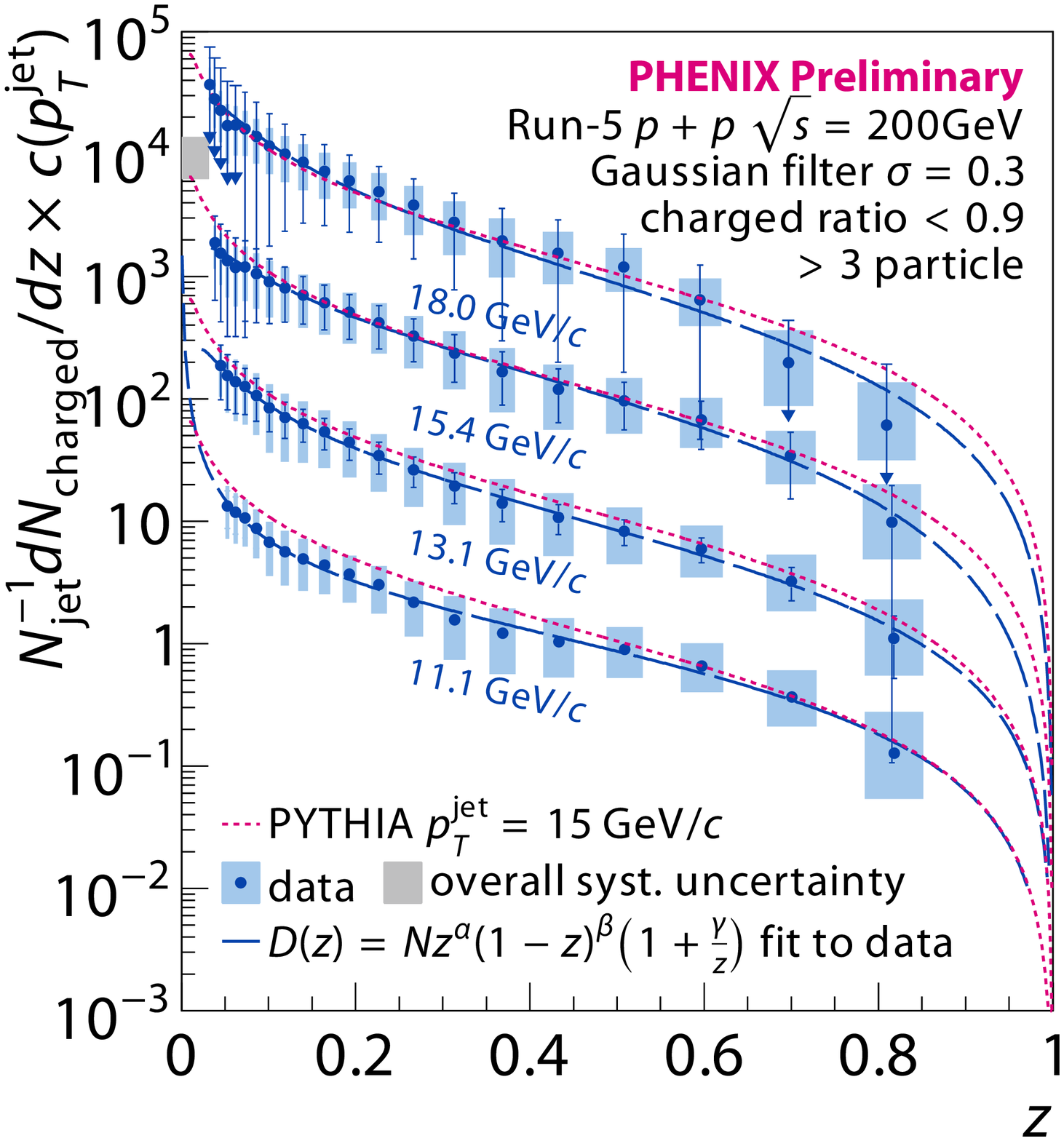}}
    \caption{PHENIX Run-5 \pp{} at \stwohundred{} charged (with
      electron rejection) jet fragmentation function with respect to
      $z = w_\mathrm{filter} p_\parallel/p^\mathrm{jet}$ and
      vertically scaled by $c(p_T^\mathrm{jet}) = 10^k, k = 0, \dots,
      3$. The shaded box to the left indicates the overall
      normalization systematic uncertainty, shaded boxes associated
      with data points indicate point-to-point systematic
      uncertainties, and error bars indicate statistical
      uncertainties. Biases from the jet level cuts are fully
      quantified in the systematic uncertainties. The blue curve
      indicates the $D(z) = N z^\alpha (1 - z)^\beta \left(1 +
        \frac{\gamma}{z}\right)$ fit to data, while the red curve
      shows the same fit from \textsc{pythia} at $p_T^\mathrm{jet} =
      15\,\mathrm{GeV}/c$.}
    \label{fig:run_5_p_p_ff_charged}
  \end{minipage}\hfill%
  \begin{minipage}[t]{3in}
    \centerline{\includegraphics[width=3in]{%
        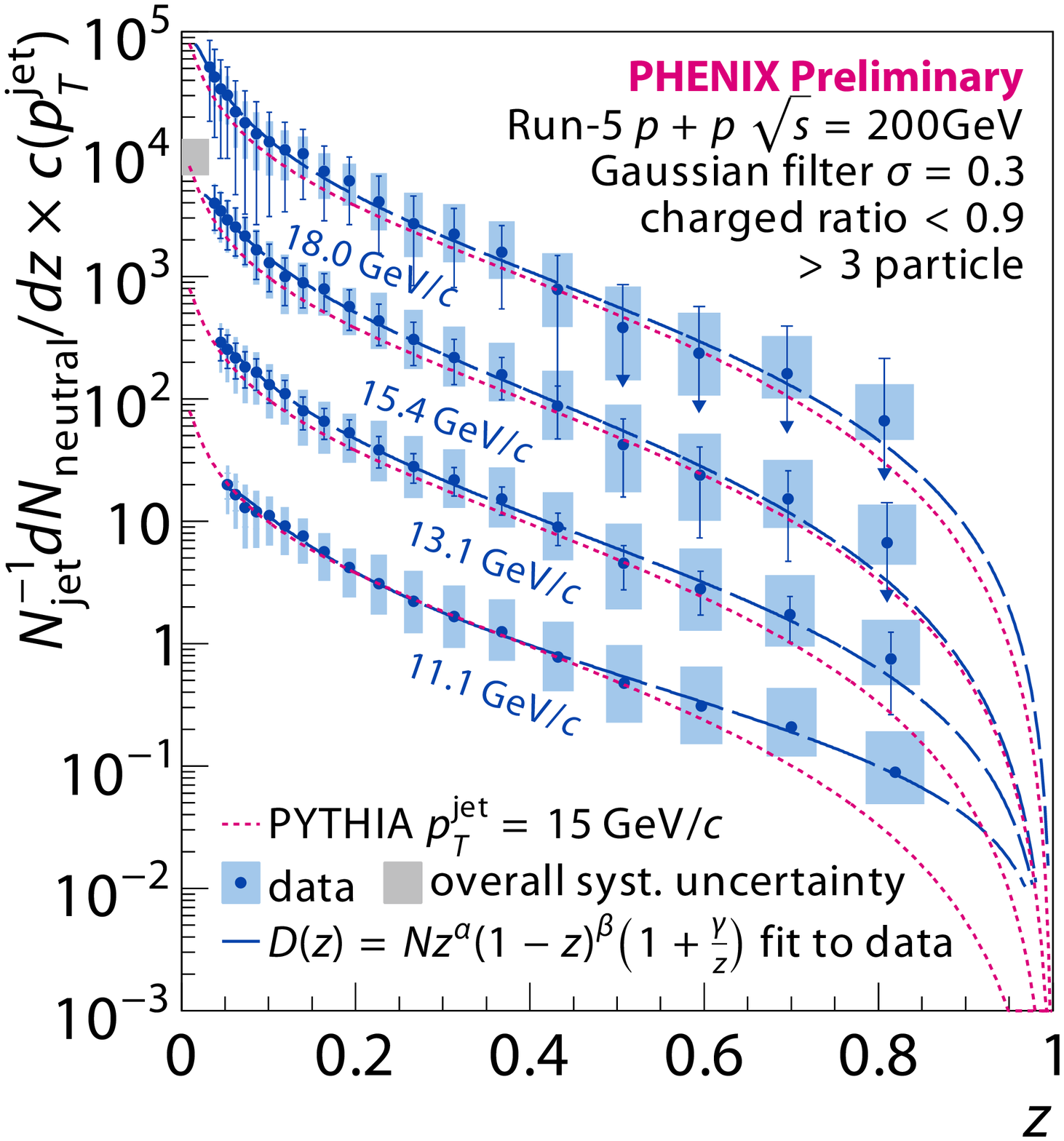}}
    \caption{PHENIX Run-5 \pp{} at \stwohundred{} (electromagnetic)
      neutral jet fragmentation function with respect to $z =
      w_\mathrm{filter} p_\parallel/p^\mathrm{jet}$ and vertically
      scaled by $c(p_T^\mathrm{jet}) = 10^k, k = 0, \dots, 3$. The
      shaded box to the left indicates the overall normalization
      systematic uncertainty, shaded boxes associated with data points
      indicate point-to-point systematic uncertainties, and error bars
      indicate statistical uncertainties. Biases from the jet level
      cuts are fully quantified in the systematic uncertainties. The
      blue curve indicates the $D(z) = N z^\alpha (1 - z)^\beta
      \left(1 + \frac{\gamma}{z}\right)$ fit to data, while the red
      curve shows the same fit from \textsc{pythia} at
      $p_T^\mathrm{jet} = 15\,\mathrm{GeV}/c$.}
    \label{fig:run_5_p_p_ff_neutral}
  \end{minipage}
\end{figure}

We use the Run-5 \pp{} minimum bias data set for the measurement of
the fragmentation function. For reconstructed jets, we measure the
longitudinal contribution of the fragment momentum to the jet as
\begin{equation}
  z = w_\mathrm{filter}(\Delta R) \frac{p_\parallel}{p^\mathrm{jet}} =
  \exp\left( -\frac{\Delta\eta^2 +
      \Delta\phi^2}{2\sigma_\mathrm{filter}} \right) \frac{\mathbf{p}
    \cdot \mathbf{p}^\mathrm{jet}}{(p^\mathrm{jet})^2}.
\end{equation}
This choice preserves the identity $\sum z = 1$ for a weighted jet
definition, compared to traditional algorithms that would assign zero
or unitary weights for nonconstituent and constituent fragments. For
practical purpose, both $p_\parallel$ and $p_T^\mathrm{jet}$ are
binned logarithmically in the $dN/(dp_\parallel dp_T^\mathrm{jet})$
distribution, which allows the $p_\parallel/p_T$ division to be
performed as bin shifts. For the narrow mid-pseudorapidity acceptance
of the PHENIX central arms, the difference in $z =
p_\parallel/p^\mathrm{jet}$ and $z = p_\parallel/p_T^\mathrm{jet}$
division is $\le 6\%$ and negligible within our uncertainties.

We use RICH to reject electrons for the measurement of charged
fragmentation function, since in PHENIX, they are produced
predominantly from $\gamma$ conversions in the beam pipe and Dalitz
decays, therefore not associated with the charged fragmentation of
jets. At the moment, we also do not reconstruct $\gamma$ pairs from
$\pi^0$ decays, although we envision to provide identified neutral
fragmentation function measurements in the future.

Since the PHENIX detector resolution for single particle far exceeds
the resolution for jet energy, we assume a diagonal transfer function
for particle energies in the transfer function
\begin{equation}
  P(p_T^\mathrm{jet,rec}, p_\parallel^\mathrm{rec} | p_T^\mathrm{jet},
  p_\parallel) = P(p_T^\mathrm{jet,rec} | p_T^\mathrm{jet}, p_\parallel)
  \delta(p_\parallel^\mathrm{rec} - p_\parallel),
\end{equation}
although the use of a full transfer function is planned for the
future. The same \textsc{geant} simulation result as for the jet
spectrum is used to evaluated the so reduced conditional transfer
function $P(p_T^\mathrm{jet,rec} | p_T^\mathrm{jet}, p_\parallel)$.

The $(p_T^\mathrm{jet}, p_\parallel)$ distribution is then unfolded by
the regularized inversion of the 2D equation
\begin{equation}
  \frac{dN}{dp_T^\mathrm{jet,rec} dp_\parallel^\mathrm{rec}} = \iint
  dp_T^\mathrm{jet} \, dp_\parallel \, P(p_T^\mathrm{jet,rec},
  p_\parallel^\mathrm{rec} | p_T^\mathrm{jet}, p_\parallel)
  \frac{dN}{dp_T^\mathrm{jet} dp_\parallel}.
  \label{eq:fredholm_2d}
\end{equation}
Similarly, the jet spectrum is unfolded by inverting
\eqref{eq:fredholm_1d}. The per-jet normalized fragmentation function
is then extracted by dividing the unfolded $(p_T^\mathrm{jet},
p_\parallel)$ distribution by the unfolded jet spectrum:
\begin{equation}
  D(z) = \frac{1}{\epsilon(z, p_T^\mathrm{jet})} \left(
    \frac{dN_\mathrm{jet}}{dp_T^\mathrm{jet}} \right)^{-1}
  \frac{dN}{dp_T^\mathrm{jet} dz} = \frac{1}{\epsilon(z,
    p_T^\mathrm{jet})} \left. \left(
      \frac{dN_\mathrm{jet}}{dp_T^\mathrm{jet}} \right)^{-1}
    p^\mathrm{jet} \, \frac{dN}{dp_T^\mathrm{jet} dp_\parallel}
  \right|_{z = p_\parallel/p^\mathrm{jet}}
\end{equation}
where $\epsilon(z, p_T^\mathrm{jet})$ is the single particle
efficiency evaluated by \textsc{geant} simulation. While the
efficiency for neutral clusters are constant at $\epsilon = 0.81\pm
0.04\text{ (syst.)}$, the efficiency for charged tracks exhibits a
strong $z$-dependence both due to the magnetic field causing low $p_T$
tracks to be bend out of the azimuthal acceptance, and tracking cuts
resulting in the rejection of high $p_T$ $\pi^\pm$ misidentified as
electrons. For $p_T^\mathrm{jet} = 18\,\mathrm{GeV}/c$, the efficiency
for charged tracks saturates for $z > 0.3$ to $\epsilon = 0.58\pm
0.06\text{ (syst.)}$.

Figure~\ref{fig:run_5_p_p_ff_charged} shows the \pp{} at
\stwohundred{} charged (with electron rejection) jet fragmentation
function, vertically scaled by $c(p_T^\mathrm{jet}) = 10^k, k = 0,
\dots, 3$. The shaded box to the left indicates the overall
normalization systematic uncertainty, shaded boxes associated with
data points indicate point-to-point systematic uncertainties, and
error bars indicate statistical uncertainties. Biases from the jet
level cuts are fully quantified in the systematic uncertainties. The
blue curve indicates the $D(z) = N z^\alpha (1 - z)^\beta \left(1 +
  \frac{\gamma}{z}\right)$ fit to data, while the red curve shows the
same fit from \textsc{pythia} at $p_T^\mathrm{jet} =
15\,\mathrm{GeV}/c$.

Figure~\ref{fig:run_5_p_p_ff_neutral} shows the same plot, for the
\pp{} at \stwohundred{} (electromagnetic) neutral jet fragmentation
function. In both set of fragmentation measurements, the fragmentation
function for $p_T^\mathrm{jet} \ge 15\,\mathrm{GeV}/c$ agree well
within our current uncertainty with \textsc{pythia}. We could reach a
maximum $z \approx 0.81$ with our measurement.

\section{Summary}

We presented the first measurement of jet fragmentation function in
\pp{} collisions performed at RHIC. It demonstrates for the ability of
PHENIX to access high-$z$ jet fragmentation property for both charged
particles and photons. Techniques to unfold multidimensional spectra
such as the fragmentation function has been discussed, which
significantly facilitated us to make such a measurement.

Restricting the data to the minimum bias set currently constrains our
statistics and ability to lower the unfolding systematic uncertainty.
Our current development of the trigger efficiency correction for
fragmentation function measurement would allow us to address this
problem in the future.

\section*{References}

\bibliography{wwnd2010-ylai}

\end{document}